\documentstyle[12pt]{article}
\title{Nucleon-Nucleon Correlations and Final State Interaction in Inclusive
Quasi-Elastic Electron Scattering off Nuclei at $x>1$}
\author{\\C. Ciofi degli Atti$^{(a)}$ and S. Simula$^{(b)}$\\\\
$^{(a)}$Department of Physics, University of Perugia and\\ Istituto Nazionale
di
Fisica  Nucleare, Sezione di Perugia,\\ Via A. Pascoli, I-06100 Perugia,
Italy\\
$^{(b)}$Istituto Nazionale di Fisica Nucleare, Sezione Sanit\'a,\\ Viale Regina
Elena 299, I-00161 Roma, Italy\\\\}
\date{}
\newcommand{\be}{\begin{eqnarray}}
\newcommand{\ee}{\end{eqnarray}}
\begin{document}
\maketitle
\abstract{ Inclusive quasi-elastic electron scattering off nuclei is
investigated
at high momentum transfer ($Q^2>1~(GeV/c)^2$) and $x>1$ adopting a consistent
treatment of nucleon-nucleon correlations in initial and final states. It is
shown that in case of light as well as complex nuclei the inclusive cross
section at $1.3<x<2$ is dominated by the absorption of the virtual photon on a
pair of correlated nucleons and by their elastic rescattering in the continuum,
whereas at $x>2$ it is governed by the rescattering of the outgoing
off-mass-shell nucleon in the complex optical  potential generated by the
ground
state of the residual (A-1)-nucleon system.\\\\\\\\}
\indent PACS number(s) : 25.30.Fj\\\\\\\\
\indent To appear in Phys. Lett. B.
\newpage
\indent Inclusive quasi-elastic (QE) electron scattering off nuclei at high
momentum transfer can provide non trivial information on the nuclear wave
function. In particular, the kinematics region corresponding to $x>1+k_F/M
\approx 1.3$ (where $k_F$ is the Fermi momentum and $x = Q^2 / 2M\nu$ the
Bjorken
scaling variable), is strongly affected by high  momentum and high removal
energy components of the nuclear wave function arising from nucleon-nucleon
(NN)
short-range and tensor correlations. Such a result has been obtained by several
authors [1-5] using the plane wave Impulse Approximation (IA). Recently,
however, it has been shown [6] that the Final State Interaction (FSI) of the
struck nucleon, evaluated with an on-shell optical potential, totally
overwhelms
the effects from ground-state NN correlations. Nonetheless, in ref. [7] the
relevance of NN correlations on the inclusive cross section in the region
$1<x<2$ has been again advocated, arguing [2] that in such a region (where
scattering from a nucleon at rest is kinematically forbidden) the contribution
from a correlated NN pair in a nucleus should be proportional to the one in the
deuteron (for which $x<2$). In order to further clarify the role played by NN
correlations and FSI, we present here the first results of a calculation of the
inclusive cross section based upon a novel approach for evaluating FSI relying
on a consistent treatment of NN correlations in initial and final states.
\newline
\indent The cross section for the inclusive process $A(e,e')X$ will be written
hereafter in the following form\\
 \be
  \sigma^{(A)} = \frac{d^2\sigma} {dE_{e'}~d\Omega_{e'}} = \sigma_0^{(A)} +
  \sigma_1^{(A)}
 \ee
\\where the contributions from different final nuclear states have been
explicitly separated out, namely $\sigma_0^{(A)}$ describes the transition to
the ground and one-hole states of the (A-1)-nucleon system and $\sigma_1^{(A)}$
the transition to more complex highly excited configurations. In what follows,
the calculation of $\sigma_0^{(A)}$ and $\sigma_1^{(A)}$, within different
levels
of sophistication for the treatment of the final A-nucleon state, will be
presented starting from the IA.\newline
\indent \bf1. The Impulse Approximation. \rm As is well known, the calculation
of
the inclusive cross section within the IA requires the knowledge of the nucleon
spectral function $P(k,E)$, which represents the joint probability to find in a
nucleus a nucleon with momentum $k \equiv | \vec{k} |$ and removal energy
$E$. In presence of ground-state NN correlations $P(k,E)$ can be written as\\
 \be
     P(k,E) = P_0(k,E) + P_1(k,E)
 \ee
\\where the subindeces 0 and 1 have the same meaning as in eq. 1, i.e. $P_0$
includes the ground and one-hole states of the (A-1)-nucleon system and $P_1$
more complex configurations (mainly 1p-2h states) which arise from 2p-2h
excitations generated in the target ground state by NN correlations. Thus, one
can write [4]
 \be
  P_0(k,E) = \frac{1} {A} \sum_{\alpha \leq F} A_{\alpha}~n_{\alpha}(k)~\delta
  \left ( E ~ - ~ | \epsilon_{\alpha} | \right )
 \ee
 where $n_\alpha$ is the nucleon momentum distribution of the single particle
state $\alpha$ having fractional occupation probability ($\int d\vec{k} ~
n_{\alpha}(k) < 1$), energy $\epsilon_\alpha$ and number of nucleons
$A_\alpha$.
As for the correlated part $P_1$, we will consider in this paper high values of
$k$ ($k>1.5~fm^{-1}$) and $E$ ($E \sim k^2/2M$), in which case, generalizing
the results of ref. [8], we get\\
 \be
   P_1(k,E) = \sum_{N_{2} = n,p} \int d \vec{k}_{cm} ~ n_{rel}^{N_1
N_2}(\vec{k}
   - {\vec{k}_{cm} \over 2}) ~ n_{cm}^{N_1 N_2}(\vec{k}_{cm}) \nonumber \\
  \delta \left [ E - E_{thr}^{(2)} - \frac{A-2} {2M(A-1)}(\vec{k} - \frac{A-1}
   {A-2}~\vec{k}_{cm})^2 \right]
\ee
\\where $n_{rel}^{N_1 N_2}$ ($n_{cm}^{N_1 N_2}$) is the momentum distribution
of
the relative (center of mass) motion of the two nucleons in a correlated pair,
and $ E_{thr}^{(2)} = M_{A-2} + 2M - M_A$ is the two-nucleon break-up
threshold.
Eq. 4 assumes that the (A-2)-nucleon system is in its ground  state, which
implies that the cm of the correlated pair moves in the mean field of the
nucleus, so that only soft components of $n_{cm}^{N_1 N_2}$ contribute to $P_1$
(see for details refs. [8-9]). Eq. 4 quantitatively reproduces the high
momentum and high removal energy parts of the many-body spectral function
calculated for $^{3}He$ [1], $^{4}He$ [10] and nuclear matter [6], as well as
the
momentum sum rule of complex nuclei [8-9].  Using eqs. 2-4 we write the IA
cross
section in the following forms, which will be useful for further elaboration\\
 \be
   [\sigma_0^{(A)}]_{IA} & = & \sum_{N=1}^A ~ \int d \vec{k} ~ dE ~ \sigma_{eN}
   ~ P_0^N (k,E) ~ \delta \left [ \nu + k^0 - E_{\vec{k} + \vec{q}} \right ]
   \\
   & = & \sum_{\alpha \leq F} ~ \int_0^\infty  dk~k~n_{\alpha} (k) ~
(Z_{\alpha}
   \sigma_{ep} + N_{\alpha} \sigma_{en}) ~ {\nu + k^0 \over q} ~ [\Delta
   (k)]_{IA}
 \ee
\\
 \be
   [\sigma_1^{(A)}]_{IA} & = & \sum_{N=1}^A ~ \int d \vec{k} ~ dE ~ \sigma_{eN}
   ~ P_1^N (k,E) ~ \delta \left [ \nu + k^0 - E_{\vec{k} + \vec{q}} \right ]
   \\
   & = & A \sigma_{Mott} \sum_{N_1 N_2 =n,p} ~ \int d \vec{k}_{cm} ~
n_{cm}^{N_1
   N_2}(\vec{k}_{cm}) ~ L^{\mu \nu} ~ [W_{\mu \nu}^{N_1 N_2}]_{IA}
 \ee
\\ In eqs. 5-8 $\nu~(\vec{q})$ is the energy (three-momentum) transfer;
$\vec{k}$
is the  momentum of the nucleon in the lab system before interaction and
$k^0 = M_A - \sqrt{(M_A + E - M)^2 + k^2}$ its off-shell energy; $E_{\vec{p}} =
\sqrt{M^2 + | \vec{p} | ^2}$; $\sigma_{eN}$ is the electron - (off-shell)
nucleon cross section; $[\Delta (k)]_{IA}$ is \\
 \be
  [\Delta (k)]_{IA} & = & \int_{E_p^-}^{E_p^+} dE_p ~ {E_p \over \nu +
  k_{\alpha}^0} ~ \delta(\nu +  k_{\alpha}^0 - E_p ) \nonumber \\
  & = & \Theta(E_p^+ - \nu - k_{\alpha}^0) ~ \Theta(\nu + k_{\alpha}^0 - E_p^-
)
 \ee
\\where $E_p^{\pm} = \sqrt{M^2 + (k \pm q)^2}$. In eq. 8 $L_{\mu \nu}$ is the
(reduced) leptonic tensor and $[W_{\mu \nu}^{N_1 N_2}]_{IA}$ the hadronic
tensor
of a correlated pair, which can be written as \\
 \be
 [W_{\mu \nu}^{N_1 N_2}]_{IA} & = & \sum_{f_{12}^0} \sum_{\beta_{12}} ~ \left [
 \langle \beta_{12} | j_{\mu}^{N_1} + j_{\mu}^{N_2} | f_{12}^0 \rangle
 \right ]^* ~ \sum_{\beta'_{12}} \left [ \langle \beta'_{12} | j_{\nu}^{N_1} +
 j_{\nu}^{N_2} | f_{12}^0 \rangle \right ] \nonumber \\
 & \cdot & \delta \left [ \nu + k_{cm}^0 - \sqrt{ (M_2^{f_{12}^0})^2 +
 (\vec{k}_{cm} + \vec{q})^2} ~ \right ]
 \ee
\\where $j_{\mu}^N$ is the nucleon current, $k_{cm}^0 = M_A - \sqrt{M_{A-2}^2 +
 | \vec{k}_{cm} | ^2}$, $| \beta_{12} \rangle$ is the relative wave
function of a correlated pair and $| f_{12}^0 \rangle$ its plane wave final
state. It should be pointed out that, whereas eqs. 5 and 7 are valid for any
spectral function, eqs. 6 and 8 results from eqs. 3 and 4, respectively. In
particular, eq. 8, whose usefulness will be clear later on, is based upon the
assumption that final and initial A-nucleon states factorize as  follows: $|
\Psi_A^f \rangle \sim \hat{A} ~ \left [ | f_{12}^0 \rangle | \vec{P}_{cm}
\rangle | \Psi_{A-2}^f \rangle \right ]$   and  $| \Psi_A^0 \rangle \sim
\hat{A} ~ \left [ | \beta_{12} \rangle | \chi_{12}^{cm} \rangle | \Psi_{A-2}^0
\rangle \right ]$ where $\hat{A}$ is a proper antisymmetrization operator, $|
\chi_{12}^{cm} \rangle$ the cm wave function of a correlated pair and $|
\vec{P}_{cm} \rangle$ its plane wave final state (note that the factorized form
for $| \Psi_A^0 \rangle$ is the basic ingredient for obtaining eq. 4 [8] and
that, using eq. 10 in eq. 8, eq. 7 is recovered in terms of the nucleon
spectral function $P_1$ given by eq. 4 with $n_{rel}^{N_1 N_2}(\vec{k}_{rel}) =
\sum_{\beta_{12}} \left | \langle \beta_{12} | \vec{k}_{rel} \rangle \right |
^2$).  The IA processes corresponding to eqs. 5 and 7 are  depicted in fig. 1a
and 1b, respectively, and the results of the calculations for $^{2}H$, $^{4}He$
and $^{56}Fe$ at $Q^2 \sim 2 ~ (GeV/c)^2$ are shown by the dotted lines in fig.
2. It can be seen that, in agreement with previous calculations [1-6], the IA
sizably underestimates the cross section at $x>1.3$; this is a very well known
fact (common to both few-body systems [1], complex nuclei [4-5] and nuclear
matter [6]) which is ascribed to the lack of any FSI within the IA (note that,
unlike the region at $x\leq1$, the effects from meson exchange  currents are
strongly suppressed at high momentum transfer and $x>1$). In what follows the
FSI will be evaluated by a novel approach based upon the observation that the
FSI involving two-nucleons emitted because of ground-state NN correlations
(process of fig. 1c, called \em two-nucleon rescattering \rm) should be
different from the FSI involving the  nucleon knocked-out from shell model
states (process of fig. 1d, called \em single nucleon rescattering \rm).
\newline
\indent \bf 2. Two-nucleon rescattering. \rm The basic assumption underlying
eq.
4 is that two nucleons are locally correlated at short separations, with their
cm being apart from the spectator (A-2)-nucleon system [2, 8]. The two-nucleon
correlation in a nucleus reminds the one acting in the deuteron; indeed, in
agreement with the results of many-body calculations [11], the nucleon momentum
distribution of complex nuclei at $k>1.5~fm^{-1}$, as predicted by eq. 4, turns
out to be the properly rescaled deuteron momentum distribution [9]. Therefore,
at high values of $k$ and $E$ the absorption of the virtual photon by a
correlated pair, which at $x>1.3$ is the dominant mechanism in the IA, is
expected to resemble the one in the deuteron; if so, such a deuteron-like
picture of the initial state should be extended also to the final state by
allowing the two nucleons to elastically rescatter as depicted in fig. 1c. An
important difference with respect to the case of the deuteron is that a
correlated pair in a nucleus is bound and moves in the field created by the
other nucleons. The details of the calculation of the two-nucleon rescattering
in the medium will be presented elsewhere [12]; here, it suffices to say that
the basic step of our approach is the replacement of the IA hadronic tensor
$[W_{\mu \nu}^{N_1N_2} ]_{IA}$ by the interacting one $W_{\mu \nu}^{N_1 N_2}$,
which is nothing but eq. 10 with the plane wave state $| f_{12}^0 \rangle$
replaced by the exact NN scattering wave function $| f_{12} \rangle$ (note that
the two-nucleon rescattering process cannot be expressed in terms of a spectral
function). It can be seen from eq. 10 that medium effects on the interacting
hadronic tensor are generated by the energy conserving $\delta$ function, in
that the intrinsic energy available to the pair is fixed by its cm
four-momentum, and, therefore, by the momentum distribution $n_{cm}^{N_1 N_2}$
appearing in eq. 8; even if the cm motion is neglected ($n_{cm}^{N_1 N_2} =
\delta(\vec{k}_{cm})$) [2], medium effects still would remain through the
quantity $k_{cm}^0$. We have calculated the inclusive cross section for the
deuteron using the RSC NN potential [13], taking into account the rescattering
in S, P and  D partial waves; then, using the same two-nucleon amplitudes
$\langle \beta_{12} | j_{\mu}^{N_1} + j_{\mu}^{N_2} | f_{12} \rangle$, we have
computed the cross section $\sigma_1^{(A)}$ for complex nuclei. The results are
shown by the  dashed lines in fig. 2: it can be seen that at $1.3<x<2$ the
process of two-nucleon rescattering brings theoretical predictions in good
agreement with experimental data. The most striking aspect of our results is
that the same mechanism which explains the deuteron data, does the same in a
complex nucleus, provided the A dependence due to $n_{cm}^{N_1 N_2}$ and
$k_{cm}^0$ (clearly exhibited in fig. 2) is properly considered. It should be
pointed out that our results hold for the whole set of kinematics considered in
refs. [14-15]. \newline
\indent \bf 3. Single nucleon rescattering. \rm It can be seen from fig. 2 that
the two-nucleon rescattering is not able to describe the experimental data at
$x>2$.  This fact is not surprising, because at $x>2$ more than two nucleons
should be involved in the scattering process. We have mocked up this process by
considering the motion of the nucleon, knocked-out from shell model states, in
the optical potential generated by the ground state of the (A-1)-nucleon system
(fig. 1d). Such an approach amounts to replace the quantity $[\Delta(k)]_{IA}$
in eq. 6 by the quantity \\
 \be
 [\Delta (k)]_{opt} = - \int_{E_p^-}^{E_p^+} dE_p ~ {E_p \over \pi (\nu +
 k_{\alpha}^0)} ~ {Im~V_{opt} \over [\nu + k_{\alpha}^0 - E_p -Re~V_{opt}]^2 +
[Im~V_{opt}]^2}
 \ee
\\ resulting from the eikonal approximation for the nucleon propagator [16]
(note
that the two-nucleon rescattering (diagram 1c) is not included in the process
described by diagram 1d, because the two-nucleon rescattering is not a multiple
scattering process, i.e., it does  not contribute to an optical potential). The
quantity $[\Delta (k)]_{opt}$ (eq. 11), calculated with the common on-shell
choice $V_{opt} = - \rho v_N \sigma_{NN} (i + \alpha_{NN} ) / 2$ ($\rho$ is the
nuclear density, $v_N$ the nucleon velocity, $\sigma_{NN}$ the total NN cross
section and $\alpha_{NN}$ the ratio of the real to the imaginary part of the
forward NN scattering amplitude) is compared in fig. 3a with $[\Delta(k)]_{IA}$
(eq. 9): it can be seen that, whereas the latter vanishes for $k<k_{min}$
($k_{min}>k_F$ at $x>1.3$), $[\Delta (k)]_{opt} \neq 0$ even when $k<k_{min}$,
so
that the low momentum part of the nucleon  momentum distribution contributes to
the cross section, resulting in a sizable increase of the low-energy tail of
the
QE peak in sharp disagreement with experimental data [17, 6]. However, treating
FSI at $x>1$ in terms of on-shell optical potentials is not justified [2, 7,
17]. Indeed, the struck nucleon, having four-momentum squared $p'^2 \cong (\nu
+
M - E)^2 - (\vec{k} + \vec{q})^2$, can be either on-mass-shell ($p'^2 = M^2$)
or off-mass-shell ($p'^2 \neq M^2$) depending on the values of $k$ and
$E$. Initial configurations with $k<k_{min}$ always give rise to intermediate
off-shell (virtual) nucleons, whose rescattering amplitudes are expected to
decrease with virtuality, for off-shell nucleons have to interact within short
times. In order to take into account off-shell effects, we have included in
$V_{opt}$ a suppression factor of the type \\
 \be
   V_{opt} = - {1 \over 2} ~ \rho v_N \sigma_{NN} (i + \alpha_{NN}) ~ e^{-
\delta
   | M^2 - p'^2 |}
 \ee
\\ where the parameter $\delta$ is the same for all kinematics considered in
this
paper. The details will be given elsewhere [12]; here, we would like to stress
that, using the suppression factor in eq. 12, ~ i) the calculated longitudinal
response of nuclear matter agrees with the "exact" one obtained in ref. [18]
within the orthogonalized correlated basis method, and ~ ii) the asymptotic
limit ($Q^2 \rightarrow \infty$) of eqs. 9 and 11 coincide, as theoretically
argued [19]. The quantity $[\Delta (k)]_{opt}$ (eq. 11) calculated using eq.
12,
is shown in fig. 3a and it can be seen that the off-shell corrections strongly
suppress the contributions from low-momentum nucleons. The effects of the
various choices of $\Delta (k)$ on the cross section $\sigma_0^{(A)}$ are shown
in fig. 3b, which clearly exhibits the damping generated by off-shell
corrections. To sum up, at $x>1$ the struck nucleon propagates in the medium
with high virtuality, so that any calculation of its rescattering in terms of
on-shell optical potentials has little to recommend itself. Our results,
including all diagrams 1a-1d, are presented in fig. 4, where the nuclear
scaling
function $F(y,q)$ (see ref. [1]) is plotted against $q^2$ for a fixed value of
the scaling variable $y$ (we remind the reader that $y=0$ ($<0$) corresponds to
$x=1$ ($>1$)). The agreement with the experimental data is good and holds in
the
whole low-energy side of the QE peak. \newline \indent In conclusion, unlike
previous approaches based upon single nucleon rescattering in the final state
[6, 17], we have also considered the two-nucleon rescattering. The main
results
of our calculations are as follows: i) at $x>2$ the inclusive cross section is
dominated by $\gamma^*$ absorption by a low momentum nucleon and by the
rescattering of the highly virtual produced nucleon in the complex optical
potential generated by the ground- state of the residual (A-1)-nucleon system;
ii) at $1.3<x<2$ the cross section is strongly affected by NN correlations,
being governed by $\gamma^*$ absorption by a pair of correlated  nucleons and
by
their rescattering in the continuum; both initial state correlations and final
state interaction resemble the ones occurring in the deuteron, apart from the
cm
motion and the binding of the pair in a complex nucleus.\\\\
We are grateful to L. Frankfurt, B.Z. Kopeliovich, T.-S. H. Lee, M. Strikman
and
D. Treleani for many enlightening discussions.  One of us (C. C.d.A.) would
like
to acknowledge G. West for his kind hospitality at LANL, where part of the
manuscript was completed.\\\\

\newpage
\begin{center}
\bf Figure Captions \rm
\end{center}

\indent Fig. 1. Processes contributing to the $A(e,e')X$ cross section: a)
one-nucleon emission within the IA; b) virtual photon absorption by a
correlated
NN pair within the IA; c) elastic two-nucleon rescattering between the emitted
nucleons of a correlated pair; d) single nucleon rescattering of a nucleon
knocked-out from shell model states.
\bigskip  \newline
\indent Fig. 2. Inclusive cross sections at $Q^2\sim2~(GeV/c)^2$ [14-15] versus
the energy transfer $\nu$. Calculations have been performed using the free
nucleon form factors of ref. [20], the cc1 prescription of ref. [21] for
$\sigma_{eN}$ and the RSC potential [13] for the NN interaction. Dotted line:
IA
(figs. 1a + 1b); dashed line: IA + two-nucleon rescattering (figs. 1a - 1c);
dot-dashed line: contribution from nucleon inelastic channels estimated as in
ref. [5].
\bigskip  \newline
\indent Fig. 3. (a) The quantity $\Delta(k)$ for the process $^{56}Fe(e,e')X$
plotted versus the nucleon momentum $k$ at $x=1.5$.  Dotted line: eq. 9 (fig.
1a); dashed line: eq. 11 (figs. 1a + 1d)  calculated using $V_{opt} = - \rho
v_N
\sigma_{NN} (i + \alpha_{NN} ) / 2$ (see text); solid line: eq. 11 (figs. 1a +
1d)  calculated using the off-shell optical potential given by eq. 12. (b) The
cross section corresponding to the three different calculations of $\Delta (k)$
shown in fig. 3a. In both figures $Q^2\sim2~(GeV/c)^2$.
\bigskip  \newline
\indent Fig. 4.  Nuclear scaling function $F(y,q)$ for $^{2}H$ (a), $^{4}He$
(b)
and $^{56}Fe$ (c) versus the squared three-momentum transfer $q^2$ for a fixed
value of the scaling variable $y=-0.4~GeV/c$ [1]. Dotted line: IA (figs. 1a +
1b); dashed line: correlated NN pair contribution (figs. 1b + 1c); dot-dashed
line: mean field contribution (figs. 1a + 1d); solid line: IA + full final
state
interaction (figs. 1a - 1d). The experimental points denoted by a full dot
correspond to the kinematics of fig. 2 and, particularly, to $\nu$ = 0.86,
0.82,
0.78 GeV for $^{2}H$, $^{4}He$ and $^{56}Fe$, respectively.

\begin{thebibliography}{99}
\bibitem{1}	C. Ciofi degli Atti, E. Pace and G. Salm\'e: Phys. Lett. \bf 127B
\rm (1983) 303; Phys. Rev. \bf C43 \rm (1991) 1155.
\bibitem{2} L.L. Frankfurt and M.I. Strikman: Phys. Rep. \bf 160 \rm (1988)
235.
\bibitem{3} X. Ji and R. McKeown: Phys. Lett. \bf 236B \rm (1990) 130.
\bibitem{4} C. Ciofi degli Atti, S. Liuti and S. Simula: Phys. Rev. \bf C41 \rm
(1990) R2474.
\bibitem{5} C. Ciofi degli Atti, D.B. Day and S. Liuti: Phys. Rev. \bf C46 \rm
(1992) 1045.
\bibitem{6} O. Benhar, A. Fabrocini, S. Fantoni, G.A. Miller, V.R.
Pandharipande
and I. Sick: Phys. Rev. \bf C44 \rm (1991) 2328.
\bibitem{7} L Frankfurt, M. Strikman, D.B. Day and M. Sargsyan: Phys. Rev.
\bf C48 \rm (1993) 2451.
\bibitem{8}	C. Ciofi degli Atti, S. Simula, L. Frankfurt and M. Strikman: Phys.
Rev. \bf C44 \rm (1991) R7.
\bibitem{9} M. Borromeo, C. Ciofi degli Atti and S. Simula: to be published.
\bibitem{10} H. Morita and T. Suzuki: Prog. Theor. Phys. \bf 86 \rm (1991) 671.
\bibitem{11} S.C. Pieper, R.B. Wiringa and V.R. Pandharipande: Phys. Rev. \bf
C46 \rm (1992) 1741.
\bibitem{12} C. Ciofi degli Atti and S. Simula: to be published.
\bibitem{13} 	R.V. Reid: Ann. Phys. (N.Y.) \bf 50 \rm (1968) 411.
\bibitem{14} W. Schutz et al.: Phys. Rev. Lett. \bf 38 \rm (1977) 259; S. Rock
et
al.: ibid. \bf 49 \rm (1982) 1139; S. Rock : private communication.
\bibitem{15}	D.B. Day et al.: Phys. Rev. Lett. \bf 59 \rm (1987) 427; Phys.
Rev.
\bf C40 \rm (1989) 1011.
\bibitem{16} S.A. Gurvitz and A.S. Rinat: Phys. Rev. \bf C35 \rm (1987) 696.
\bibitem{17} T. Uchiyama, A.E.L. Dieperink and O. Scholten: Phys. Lett. \bf
233B
\rm (1989) 31.
\bibitem{18} A. Fabrocini and S. Fantoni: Nucl. Phys. \bf A503 \rm (1989) 375.
\bibitem{19} S.A. Gurvitz, A.S. Rinat and R. Rosenfelder: Phys. Rev. \bf C40
\rm
(1989) 1363; E. Pace, G. Salm\'e and G.B. West: Phys. Lett. \bf 273B \rm (1991)
205. \bibitem{20}	S. Galster et al.: Nucl. Phys. \bf B32 \rm (1971) 221.
\bibitem{21} T. De Forest: Nucl. Phys. \bf A392 \rm (1983) 232.
\end{thebibliography}
\end{document}